\documentclass{statsoc}

\usepackage{geometry}
\usepackage{graphicx,psfrag,epsf}
\usepackage{enumerate}
\usepackage{mathtools}
\usepackage{natbib}
\usepackage[hyphens]{url} % not crucial - just used below for the URL 
\usepackage{hyperref}
\usepackage{floatrow}
\usepackage{float}
\usepackage{bm}
\usepackage[title]{appendix}
\usepackage{ulem}

\title[Global Household Energy Model]{Global Household Energy Model: A Multivariate Hierarchical Approach to Estimating Trends in the Use of Polluting and Clean Fuels for Cooking}

\author[Stoner {\it et al.}]{Oliver Stoner, Gavin Shaddick and Theo Economou}
\address{Department of Mathematics, University of Exeter,
Exeter,
UK}
\author[Stoner {\it et al.}]{Sophie Gumy, Jessica Lewis, Itzel Lucio, Giulia Ruggeri and Heather Adair-Rohani}
\address{World Health Organization, Geneva, Switzerland}

\begin{document}
\begin{abstract}
In 2017 an estimated 3 billion people used polluting fuels and technologies as their primary cooking solution, with 3.8 million deaths annually attributed to household exposure to the resulting fine particulate matter air pollution. Currently, health burdens are calculated using aggregations of fuel types, e.g. solid fuels, as country-level estimates of the use of specific fuel types, e.g. wood and charcoal, are unavailable. To expand the knowledge base about impacts of household air pollution on health, we develop and implement a Bayesian hierarchical model, based on Generalized Dirichlet Multinomial distributions, that jointly estimates non-linear trends in the use of eight key fuel types, overcoming several data-specific challenges including missing or combined fuel use values. We assess model fit using within-sample predictive analysis and an out-of-sample prediction experiment to evaluate the model's forecasting performance.
\end{abstract}

\section{Introduction}
\label{sec:intro}
In 2017, an estimated 3 billion people, or 39\% of the global population, used a solid fuel (charcoal, coal, crop residues, dung, or wood) or kerosene as their primary fuel for cooking. This results in the emission of  dangerous levels of pollutants, including fine particulate matter (PM$_{2.5}$) and carbon monoxide \citep{guidelines}. The World Health Organization (WHO) has estimated that about 3.8 million deaths per year worldwide can be attributed to pollution from household cooking \citep{pressrelease}. This harm is  compounded by  the burden on people, notably women and children, who must dedicate large amounts of time to fuel collection which might otherwise be spent on education or work, and the risk of burn injuries.

To address this leading cause of disease and premature death in low- and middle-income countries, the 2030 Agenda for Sustainable Development, adopted by all United Nations member states, set a target of universal access to clean fuels and technologies for cooking (Sustainable Development Goal (SDG) 7.1.2) and to substantially reduce the number of deaths from the joint effects of ambient and household air pollution (SDG 3.9). Although there have been improvements in the proportion with access to clean fuels and technologies in some regions,  globally  these have been largely outpaced by population growth. This means that the absolute number of people without access to clean fuels and technologies has stagnated, decreasing only by 3\% between 2000 and 2017. As a result, the world is only projected to achieve 74\% clean fuel use by 2030 under current policy scenarios \citep{trackingreport}.

In 2016 the World Health Assembly adopted a roadmap consisting of four priority areas of action to tackle the health risks of air pollution, notably \textit{`expanding the knowledge base about impacts of air pollution on health'} \citep{roadmap}. Currently, the WHO publishes estimates of `polluting fuel use' and `clean fuel and technology use', representing the combined use of all polluting fuels and all clean fuels and technologies, respectively, for SDG monitoring. Here `use' is defined as the proportion of people primarily relying on a given fuel or technology for cooking. In addition, the WHO presently assumes that `clean fuel use' = `clean fuel and technology use', due to the limited availability of data on the types of stoves used for cooking and the current absence of any scalable biomass stoves which can be considered `clean' for health. These estimates are available for most countries, separately for urban and rural areas where fuel use trends often differ systematically, and for each year between 1990 and 2017. Conventionally, these estimates then serve as a practical surrogate for estimating the global burden of disease associated with using polluting fuels for cooking \citep{bonjour2013solid}. However, basing estimates of health impacts on the combined use of polluting fuels  fails to take into account variation in the risks associated with different fuels and technologies. Recently, \cite{SHUPLER2018354} introduced a method for estimating exposure for several specific fuel types that takes into account variation in exposure between countries. Despite this, global burden of disease estimates based on the use of specific fuels remain unavailable, as this would also require global estimates of specific fuel use.

In this article, by developing and implementing  a model for the use of eight specific fuel types, we make a substantial contribution to  the expansion of the knowledge base on the impacts of household air pollution.  Our aims are to:
\begin{itemize}
\item[(i)] Estimate trends in specific fuel usage, together with coherent estimates of uncertainty.
\item[(iii)] Provide meaningful estimates of individual fuel usage for countries where data is limited.
\item[(iiii)] Predict present-day fuel usage, addressing lags in data collection, and project estimated trends into the future.
\end{itemize}

Trends in the use of specific fuel types are modelled together with  survey sampling variability, which may vary between urban and rural areas and by country. Where data for a given country is limited, the model structure can derive information from regional trends. The model allows for different fuel use trends in urban and rural areas and is able to produce predictions (with associated uncertainty) of future use of different fuel types, providing policy makers with a baseline against which they can evaluate the effectiveness of future interventions. 

The remainder of the paper is organized as follows: Section \ref{sec:method} provides details of the available data and the proposed modelling approach, including the implementation of the model using Markov chain Monte Carlo (MCMC); Section \ref{sec:valid} presents posterior predictive model checking and a future forecasting experiment; and, finally, Section \ref{sec:conclusion} provides an overall summary and a concluding discussion of the model's impact. 

\section{Methodology}
\label{sec:method}
Information on the types of technologies and fuels used by households for cooking is regularly collected in nationally-representative household surveys or censuses and compiled in the WHO Household Energy Database \citep{database}. As of mid 2019, the database contains over 1100 surveys, with over 150 countries having at least one survey over the period 1990 to 2017. For each survey, the database contains the proportion of surveyed households using as their primary cooking fuel each of 10 key types: biogas; charcoal; coal; crop residues; dung; electricity; kerosene; liquid petroleum gas (LPG); natural gas; and wood.

\begin{figure}[h!]
\floatbox[{\capbeside\thisfloatsetup{capbesideposition={right,center},capbesidewidth=0.4\linewidth}}]{figure}[1.15\FBwidth]
{\caption{Smooth density estimates of the proportion of survey respondents relying on wood as their primary cooking fuel by WHO region, from all surveys contained in the WHO Household Energy Database (1990-2017).} \label{fig:region}}
{\includegraphics[width=1\linewidth]{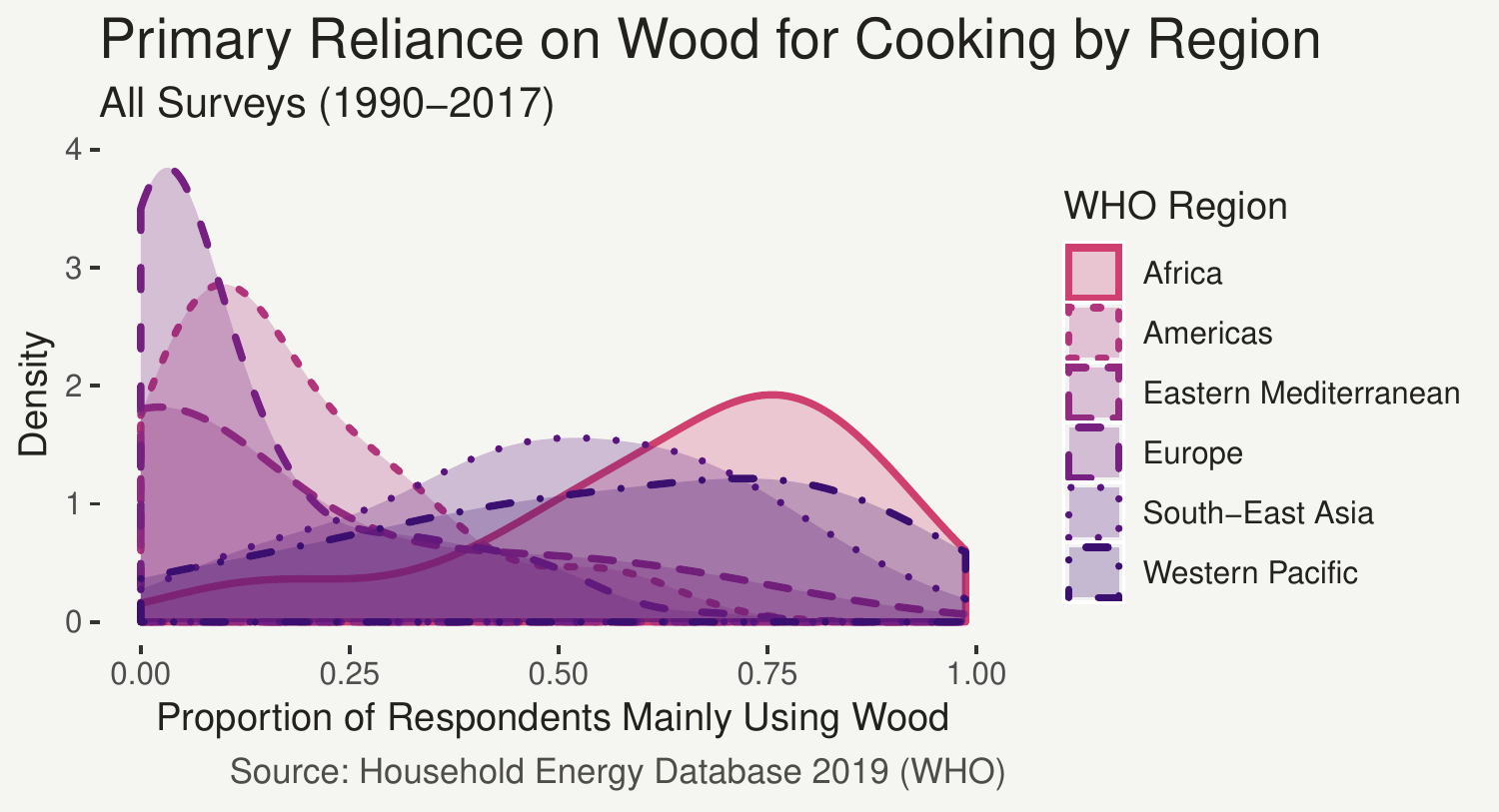}}
\end{figure}

Over the period 1990 to 2017, the average number of surveys per country per year is around 0.3. Even if survey coverage were far greater, survey sampling variability means that individual surveys would still not be a reliable indicator on which to base policy decisions. Statistical models can be used to separate trends from sampling variability, while also allowing uncertainty in the trends to be appropriately quantified. Information from other sources, such as economic or social indicators, can also be included to allow for more reliable inference in countries with few surveys. For example, \cite{Rehfuess2006} use regression methods to quantify the association between solid fuel usage and a number of socio-economic factors to predict usage in countries where no data was available. An alternative source of information which can be exploited by statistical models is that the proportion of people using each fuel type as their primary cooking fuel tends to be more similar, on average, between countries in the same region, than between countries in different regions. Figure \ref{fig:region} illustrates differences in wood use by WHO region, with  smooth density estimates of the proportion of households using wood as their primary cooking fuel, from surveys in years 1990 to 2010. For example, the density estimates suggest that use of wood is more prevalent in African countries than in European countries over this period. 

Using this data to evaluate trends in the use of specific fuels presents a number of challenges related to inconsistencies in both the quality and quantity of information that is available from the surveys. We specifically address four of these issues in our modelling approach:

\begin{itemize}
	\item[(a)] Many surveys report fuel values which are in some sense incomplete. This often includes combining more than one specific fuel type (e.g. LPG and natural gas) into a single option in the survey (e.g. gas). In some cases this can arise due to  cultures and/or languages having a single term which includes several distinct fuel types (e.g. the French language term `charbon' which can include both coal and charcoal). Another common problem is inconsistency in how sub-fuels are categorised: for example, grass may be included in the crop residues category in one survey and in the dung category in another. Other, less common, issues include  non-exhaustive lists of individual fuel options, with key fuels included  in  an `other' category, resulting in missing values for those fuels. These issues mean that the time series of survey values  for some fuels in some countries  can be highly unstable. 
	\item[(b)] The total number of respondents is only available  for approximately 50\% of surveys in the database. For surveys where this information is not available, only the proportions using each fuel are given and the original counts (the number of respondents using each fuel) are non-recoverable. 
	\item[(c)] Information on trends in the use of specific fuels  is required for both urban and rural areas but, in many cases,  surveys only provide data for the overall population. 
	
	%\item[(d)] The urban-rural stratification used within some surveys may not be consistent with the proportions (of urban and rural populations) suggested by UN estimates. If systematic within a country, this has the potential to bias  estimates. 
	
\end{itemize}

\subsection{Generalised-Dirichlet-Multinomial}
\label{sec:GDM}
For clarity of exposition, the following explanation relates to $y_i$, the number of respondents in a survey using fuel type $i$ as their primary fuel for cooking,  ignoring for now any indices related to the country and the year. If we knew the total number of survey respondents $n$ for all data, a first approach to modelling could be to assume that data on $\bm{y}=\{y_i\}$ arise from a $\mbox{Multinomial}(\bm{p},n)$ distribution. Then $p_i$ would represent the proportion of people in the population using fuel $i$. This assumes that the survey sample is representative of the overall population. In reality, survey samples are imperfect and the Multinomial model may not be sufficiently flexible to capture the extra variability caused by flaws in the survey design. For instance, the survey may not cover the whole geographical area of interest.

A flexible extension of this approach is to model $\bm{y}$ using a Generalized Dirichlet Multinomial$(\bm{\alpha},\bm{\beta},n)$ (GDM) distribution, a mixture of the Generalized Dirichlet (GD) with probability density function (pdf): 
\begin{equation}
p(p_1,p_2,...,p_k\mid \bm{\alpha},\bm{\beta}) =  p_k^{\beta_{k-1}-1} \prod_{i=1}^{k-1} \left[ \frac{p_i^{\alpha_i-1}}{B(\alpha_i,\beta_i)}   \left(\sum_{j=i}^k p_j \right)^{\beta_{i-1}-(\alpha_i+\beta_i)} \right]  \label{GD}  
\end{equation}
and the Multinomial distribution, so that 
\begin{equation}
\bm{p} \sim \mbox{Generalized-Dirichlet}(\bm{\alpha},\bm{\beta}); \qquad \bm{y} \mid \bm{p} \sim \mbox{Multinomial}(\bm{p},n).
\end{equation}
The marginal probability mass function (pmf) of the GDM is then:
\begin{equation}
p(y_1,y_2,...,y_k\mid \bm{\alpha},\bm{\beta},n) = \frac{\Gamma(n+1)}{\Gamma(y_k+1)}\prod_{i=1}^{k-1} 
\left[ \frac{\Gamma(y_i+\alpha_i)\Gamma(\sum_{j=i+1}^k y_j + \beta_i)}{B(\alpha_i,\beta_i)\Gamma(y_i+1)\Gamma(\alpha_i+\beta_i+\sum_{j=i}^k y_j)} \right]. \label{GDM}
\end{equation}
Any additional variability caused by non-representative sampling can be potentially captured by the GD component. The GD also has a very flexible covariance structure compared to the Dirichlet, which it reduces to in the special case that $\beta_i=\alpha_{i+1}+\beta_{i+1}$ for $i \in 1,...,k-2$ and $\beta_{k-1}=\alpha_k$.

Recall from Section \ref{sec:method} (b) that, for around half of the available data, only the proportion $\bm{x}=\{y_i/n\}$ of respondents using each fuel is available, with the total number of respondents $n$ being unknown. This means that we cannot use the GDM to directly model the number of respondents primarily using each fuel, if we wish to use all of the available data. However, as the principal interest lies in estimating or predicting trends the fuel usage proportions $\bm{x}$, an alternative approach would be to model the proportions themselves, for example using a GD distribution. In that case, though, the presence of many 0\% and 100\% fuel usage observations (which fall outside the range space of the GD) make this impractical. Instead, we opt for an approximate procedure for modelling $\bm{x}$, namely by transforming observations of $x_i$ into conceptual counts $v_i$, out of a chosen total $N$. To ensure that the sum of the transformed counts does not exceed $N$, one can compute $v_i = \lfloor N x_i \rfloor$ (using the floor function, as opposed to rounding). The counts $\bm{v}$ can then be modelled as GDM$(\bm{\alpha},\bm{\beta},N)$, so that predictions are based on $v_i/N$. The idea behind this is that the flexibility of the GDM means that we can still capture the distribution of $\bm{x}$ well:  any variability lost or gained from the Multinomial component, by respectively using a larger or a smaller $N$ compared to the original $n$, can be accounted for by appropriate adjustment in the parameters of the GD component.

In Appendix \ref{app:simulation}, we present a simulation study using the observed sample sizes $n$ from the data. We illustrate that this approximate method yields an inference for the population-wide fuel usage which converges (as $N$ increases) to the inference obtained by modelling $\bm{y}$ directly. Our simulation experiment suggested that values greater than $N=10000$ are likely sufficiently large, so we conservatively opt for $N=100000$. This results in a virtually zero contribution to the variability of $\bm{v}/N$ from the Multinomial component, bearing that in mind that the GD component can absorb any additional variation associated with smaller sample sizes.

\subsection{Tiered approach}\label{sec:tiers}
To motivate the way in which we will employ the GDM for these data, it is instructive to consider Figure 2, which illustrates key cooking fuel types and how they are typically aggregated into more general classifications, e.g. solid fuels. In principle, it is possible to model the use of specific fuels directly using the GDM:
\begin{eqnarray}
v_1,\dots,v_{11} &\sim & \mbox{GDM}(\bm{\alpha},\bm{\beta},N); \label{eq:individual_gdm}\\
\{1,\dots, 11\} &\equiv &\{{\mbox{wood}},{\mbox{cropwaste}},{\mbox{dung}},{\mbox{charcoal}},{\mbox{coal}},{\mbox{kerosene}},\\&&{\mbox{electricity}},{\mbox{LPG}},{\mbox{natural gas}},{\mbox{biogas}},{\mbox{others}}\}.\nonumber
\end{eqnarray}
Predictions for aggregate groups, e.g. solid fuels, can then be achieved by aggregating predictions for the individual fuels. However, recall that one of the key challenges with modelling this data, (a), is inconsistency in data collection. For example, some surveys combine more than one fuel type (e.g. charcoal and coal) into a single category. Furthermore, there is sometimes inconsistency in the way surveys categorise sub-fuels (e.g. grass). The result of this issue is that, for some countries, the time series of affected individual fuels are unstable. As such, modelling the use of all individual fuel types with one GDM (as in \eqref{eq:individual_gdm}) will adversely impact estimates for the mean trends, sampling variability and any associated uncertainty, not just for affected fuels but for the other fuels as well, owing to the multivariate nature of the model and the data.
\begin{figure}[h!]
\floatbox[{\capbeside\thisfloatsetup{capbesideposition={left,center},capbesidewidth=0.4\linewidth}}]{figure}[1.15\FBwidth]
{\caption{Hierarchy of cooking fuel types in the Global Household Energy Model.} \label{fig:tree}}
{\includegraphics[width=1\linewidth]{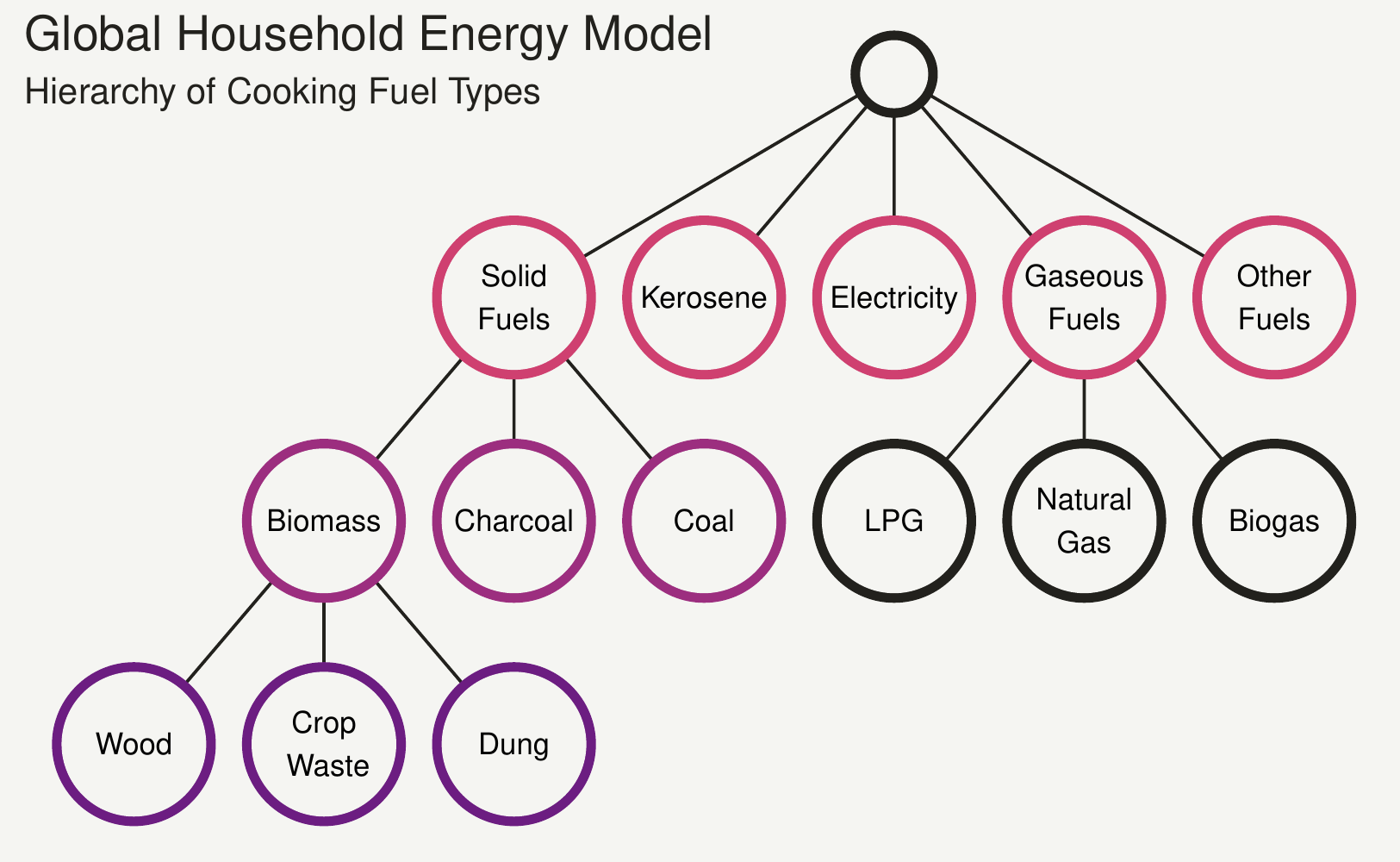}}
\end{figure}

Fortunately, as they are the result of `confusion' among certain fuel types, these issues can be resolved by aggregating individual fuels into more general fuel types. For example, confusion between wood, cropwaste and dung can be resolved by aggregating data for these fuels into the more general category `biomass' (which in this paper includes raw/unprocessed biomass fuels but excludes charcoal), while any outstanding confusion between charcoal and coal or between charcoal and wood can be resolved by aggregating into `solid fuels'. Similarly, LPG and natural gas are very commonly combined at the survey level, which can be recognised by the formation of a `gas' aggregate category.

This motivates the adoption of a tiered approach, where the use of the most aggregated fuel categories (e.g. solid fuels and gaseous fuels) are modelled as GDM at the `top' tier (note that the tier does not relate to the merits or abundance of each fuel, only how we organise the fuels for modelling purposes), alongside other fuels that are unlikely to be confused or combined (e.g. kerosene and electricity) and an aggregation of other minor fuels and technologies (e.g. alcohol, solar stoves):
\begin{eqnarray}
\{v_{solid},v_{kerosene},v_{gas},v_{electricity},v_{others}\}&\sim & \mbox{GDM}(\bm{\alpha},\bm{\beta},N).\label{eq:top}
\end{eqnarray}
This ensures that any instabilities arising from erroneous convolution of individual fuel types, e.g. charcoal and coal, does not propagate into the other fuel categories in the top tier. These categories can then be progressively disaggregated through nested GDM models. As in some countries there is convolution between biomass fuel types (e.g. wood and cropwaste), fully disaggregating solid fuels means that, in these countries, predictions for charcoal and coal will still be needlessly impacted. To address this, a `mid' tier is introduced to aggregate the biomass fuel types and model these alongside charcoal and coal:
\begin{eqnarray}
\{v_{biomass},v_{charcoal},v_{coal}\}&\sim & \mbox{GDM}(\bm{\alpha},\bm{\beta},v_{solid}).\label{eq:mid}
\end{eqnarray}
The biomass fuel types can then be disaggregated in the `lower' tier with a third GDM model:
\begin{eqnarray}
\{v_{wood},v_{cropwaste},v_{dung}\}&\sim & \mbox{GDM}(\bm{\alpha},\bm{\beta},v_{biomass}). \label{eq:lower}
\end{eqnarray}
We could then disaggregate `gas' into the three individual gaseous fuels with a fourth GDM model (a parallel mid-tier). This is however not essential for our application (estimating population exposure to household air pollution) as the difference between the different gaseous fuels in terms of pollutant concentrations is minimal compared to the difference between the gaseous fuels and the polluting fuels \citep{guidelines}. Following this approach, the result is that a joint predictive inference for 8 individual fuel types is achieved, but in a way which prevents inconsistency in particular fuel types from affecting the others. 

\subsection{Conditional models}\label{sec:conditional}
Recall that an additional challenge, (a), is that occasionally a value $x_i$ (and thus $v_i$) is missing for at least one individual fuel (for a given country-year combination). To model this data in a way that easily allows prediction of the missing fuel values, we implement each GDM (from the three tiers) using the implicit conditional densities rather than the joint one. Specifically, for counts $\bm{v}$ and total $N$, the conditional distribution of (fuel) $v_i$ given the others is:
\begin{eqnarray}
v_i\mid v_{-i}, \bm{\alpha}, \bm{\beta} &\sim & \mbox{Beta-Binomial}\left(\alpha_i,\beta_i,n_i=N-\sum_{j<i} v_j\right)  \\
p(v_i\mid v_{-i}, \bm{\alpha}, \bm{\beta}) &=& \binom{n_i}{v_i} \frac{B(v_i+\alpha_i ,n_i-v_i+\beta_i)}{B(\alpha_i , \beta_i)}
\end{eqnarray}
Fitting this model in a Bayesian setting implies that any missing values $v_i$ can be sampled using Markov Chain Monte Carlo (MCMC). Furthermore, for ease of interpretation we re-parameterize the conditional distributions in terms of their expectations $\nu_i$ and dispersion parameters $\phi_i$: 
\begin{equation}
\alpha_i = \nu_i \phi_i; \qquad \beta_i = (1-\nu_i) \phi_i 
\end{equation}
The relative mean $\nu_i$ is interpreted as the expected proportion of households using fuel $i$ out of those not using any of the fuels higher up the hierarchy $(1,\ldots,i-1)$. For example, $\nu_1$ is the expected proportion who use wood from the whole population, $\nu_2$ is the proportion who use charcoal from the population who do not use wood and $\nu_3$ is the proportion who use coal from the population who use neither wood nor charcoal. Through parameter $\phi_i$, the model is able to compensate for any gain or loss of variance in the conditional Multinomial model for $v_i$ caused by the introduction of the ``artificial'' total $N$. For more interpretable inference, the marginal mean vector of proportions $\bm{\mu}=\{\mu_i\}$ of households relying on each fuel $i$ can be recovered from the relative means $\nu_i$:
\begin{align}
\mu_1 = \nu_1;\qquad  \mu_k = \nu_k\prod_{i=1}^{k-1}(1-\nu_i)\qquad k\geq 2. \label{muK}
\end{align}

\subsection{Country and regional models}
Introducing indices for a survey conducted in area $j$ (1=urban, 2=rural) of country $c$ and in year $t$, the characterisation of the relative mean $\nu_{i,j,c,t}$ is defined by:
\begin{equation}\label{eq:nu}
\log \left(\frac{\nu_{i,j,c,t}}{1-\nu_{i,j,c,t}} \right) = f_{i,j,c}(t),
\end{equation}
where the logistic transformation ensures that $\nu_{i,j,c,t}\in (0,1)$. Here we characterise  functions $f$ as linear combinations of an intercept term, a linear term and non-linear thin-plate spline terms:
\begin{eqnarray}
f_{i,j,c}(t) = \beta_{0,i,j,c} + \beta_{1,i,j,c} X_{t,1} + \sum_{k=2}^{K}\beta_{k,i,j,c} X_{t,k}. 
\end{eqnarray}
Here $\bm{X}$ is a model matrix of spline terms, where $X_{t,1}$ is linear in time and $X_{t,2},\dots,X_{t,K}$ are non-linear thin-plate terms. The choice of the number of basis terms $K$ must be made a-priori, which corresponds to an upper bound on flexibility (similar to choosing a number of polynomial terms). For larger $K$ the functions are penalised for smoothness parametrically. Here we choose $K=10$, which is approximately one basis term for every three years.  All coefficients are modelled as random effects, whose prior distributions have expectations (characterised as fixed effects) that vary with the region the country is in (denoted by region index $r(c)$). Several choices are available for regional classifications, such as the 6 WHO regions, the 21 Global Burden of Disease (GBD) regions and the 7 GBD super-regions \citep{DIMAQ}, or the 8 SDG regions. For a chosen regional classification, the country random effects are modelled as:
\begin{eqnarray}
\beta_{0,i,j,c} &\sim & \mbox{Normal}(\gamma_{0,i,j,r(c))},\sigma_{0,i,j}^{2(\beta)}); \label{eq:intercept}\\
\beta_{1,i,j,c} &\sim & \mbox{Normal}(\gamma_{1,i,j,r(c)},\sigma_{1,i,j}^{2(\beta)}); \\
\{\beta_2,\dots,\beta_K\} & \sim & \mbox{Multivariate-Normal}(\{\gamma_2,\dots,\gamma_K\},\bm{\Omega}_{i,j,c}^{-1(\beta)}). \label{eq:country_spline}
\end{eqnarray}	
The regional parameters (e.g. $\gamma_{0,i,j,r(c)}$) are then modelled as thin-plate spline (fixed-effect) coefficients. It can be shown that this model is equivalent to the additive combination of a regional-level spline and a country-level spline, where the former is a mean trend while the latter captures country-level deviation from this mean. The advantage of this characterisation over explicitly separating the country and regional effects into two different splines, however, is improved MCMC efficiency. Each precision matrix $\bm{\Omega}_{i,j,c}^{-1(\beta)}$ is known  \citep{jagam}, and scaled by parameter $\lambda_{i,j,c}^{(\beta)}$ which penalizes the thin-plate spline (specifically the deviation of the country spline from the regional spline) for smoothness, to avoid over-fitting. Each $\log(\lambda_{i,j,c}^{(\beta)})$ is modelled as a random effect arising from a Normal($\upsilon_{i,j}^{(\beta)}$,$\sigma_{2,i,j}^{2(\beta)}$) prior distribution.

The purpose of treating the coefficients $\bm{\beta}_{i,j,c}$ and smoothing parameters $\lambda_{i,j,c}^{(\beta)}$ as random effects is to improve prediction in countries with sparse data, where regional trends and global hyper-parameters can constrain the overall country effect to not be too extreme with respect to other countries in the same region. We trialled this approach using the 6 WHO regions and alternatively the 21 GBD regions. Using the 6 WHO regions, we found they encompassed too broad a range of fuel use patterns to be particularly useful for improving prediction in countries with little data. Conversely, when using the 21 GBD regions we found that they often contained too few countries, or had too many countries with little data, to allow the precise estimation of regional trends.

To address the issues posed by this choice, we opted for a nested model which utilises both GBD regional structures and GBD super-regional structures (denoted by index $s(r)$):
\begin{eqnarray}
\gamma_{0,i,j,r} &\sim & \mbox{Normal}(\theta_{0,i,j,s(r))},\sigma_{0,i,j}^{2(\gamma)}) \label{eq:region_intercept}\\
\gamma_{1,i,j,r} &\sim & \mbox{Normal}(\theta_{1,i,j,s(r)},\sigma_{1,i,j}^{2(\gamma)}) \\
\{\gamma_2,\dots,\gamma_K\} & \sim & \mbox{Multivariate-Normal}(\{\theta_2,\dots,\theta_K\},\bm{\Omega}_{i,j,r}^{-1(\gamma)}) \label{eq:region_spline}
\end{eqnarray}
The regional thin-plate spline coefficients (e.g. $\gamma_{0,i,j,r(c)}$) are now also modelled as random effects, with super-regional expectations. Each precision matrix $\bm{\Omega}_{i,j,r}^{-1(\gamma)}$ is, as before, a known matrix scaled by parameter $\lambda_{i,j,r}^{(\gamma)}$, to penalize for smoothness (of the deviation of the regional trend from the super-regional trend). The penalty parameters $\lambda_{i,j,r}^{(\gamma)}$ are once more modelled at the log-scale as random effects, arising from a Normal($\upsilon_{i,j}^{(\gamma)}$,$\sigma_{2,i,j}^{2(\gamma)}$) prior distribution. The super-regional intercept and linear terms are then modelled as fixed effects:
\begin{eqnarray}
\theta_{0,i,j,s} &\sim & \mbox{Normal}(0,10^2); \label{eq:super_region_intercept}\\
\theta_{1,i,j,s} &\sim & \mbox{Normal}(0,10^2); \\
\{\theta_2,\dots,\theta_K\} & \sim & \mbox{Multivariate-Normal}(\bm{0},\bm{\Omega}_{i,j,s}^{-1(\theta)}). \label{eq:super_region_spline}
\end{eqnarray}
Each $\bm{\Omega}_{i,j,s}^{-1(\theta)}$ is scaled by the fixed effect $\lambda_{i,j,s}^{(\theta)}$, whose prior distribution is Normal($0,10^2$) at the log-scale. Now each $f_{i,j,c}(t)$ is equivalent to the additive combination of a super-regional spline, a regional deviation  spline, and a country deviation spline. By adopting a nested regional structure, countries with little data benefit from more precise regional trend estimation, borrowing information from other countries in the same GBD region that have sufficient data. Failing that, further borrowing is achieved from the super-regional trend.

Having specified models for the relative mean proportions $\nu_{i,j,c,t}$, it remains to define a model the dispersion parameters $\phi_{i,c}$. Recall that these parameters are intended to capture additional survey variability (compared to the Multinomial), which is affected by the introduction of an `artificial' sample size $N$.  In countries with little data, we would like to constrain survey variability to reasonable levels, so we $\phi_{i,c}$ as random effects which vary by country:
\begin{eqnarray}
\log(\phi_{i,c}) \sim \mbox{Normal}(\upsilon_{i,j}^{(\phi)},\sigma_{i,j}^{2(\phi)}).
\end{eqnarray}
In the absence of any belief that survey variability should be regionally structured, the random effects are constrained by global hyper-parameters $\upsilon_{i,j}^{(\phi)}$ and $\sigma_{i,j}^{2(\phi)}$.

\subsection{Urban and rural variability}
A further challenge with modelling this data, (c), is that while most surveys in the data report both urban and rural values, some only report an overall value for the whole sample. So that these surveys can inform the estimation of urban and rural trends, we incorporate a layer in the model which relates the marginal mean proportions of urban, rural, and overall values as follows:
\begin{eqnarray}
\bm{\mu}_{c,t}^{overall} &=& \pi_{c,t}\bm{\mu}_{c,t}^{urban} + (1-\pi_{c,t})\bm{\mu}_{c,t}^{rural}; \label{MUoverall}\\
\log{\left(\frac{\pi_{c,t}}{1-\pi_{c,t}}\right)} &=& \log{\left(\frac{P_{c,t}}{1-P_{c,t}}\right)} +
g_c (t). \label{eq:urban_proportion}
\end{eqnarray}
The overall mean fuel usage proportions $\bm{\mu}_{c,t}^{overall}$ are then defined as a weighted sum in \eqref{MUoverall}, of the mean rural and urban proportions. The weights $\pi_{c,t}\in(0,1)$ represent the mean proportion of survey respondents living in an urban area, in country $c$ and year $t$. To capture structured demographic variability between countries and over time, UN estimates \citep{WUP} of the proportion of people living in an urban area for each country and year, $P_{c,t}$, are included as offsets in the model for $\pi_{c,t}$. For each country, any remaining structured variability in the urban proportion is modelled using a smooth function $g_c(t)$. These functions should ideally be flexible enough to capture the mean urban proportions well. However, from a modelling perspective, they also introduce extra degrees of freedom to capture the overall survey observations well. Therefore, to avoid over-fitting, we once again employ penalized thin-plate splines for $g_c(t)$:
\begin{eqnarray}
g_{c}(t) &=& \kappa_{0,c} + \kappa_{1,c} X_{t,1} + \sum_{k=2}^{K}\kappa_{k,c} X_{t,k}; \\ 
\kappa_{0,c} &\sim & \mbox{Normal}(0,\sigma_{0}^{2(\kappa)}); \label{eq:urban_intercept}\\
\kappa_{1,c} &\sim & \mbox{Normal}(0,\sigma_{1}^{2(\kappa)}); \\
\{\kappa_2,\dots,\kappa_k\} & \sim & \mbox{Multivariate-Normal}(\{\kappa_2,\dots,\kappa_k\},\bm{\Omega}_{c}^{-1(\kappa)}). \label{eq:urban_spline}
\end{eqnarray}
Each precision matrix $\bm{\Omega}_{c}^{-1(\kappa)}$ is, for one final time, a known matrix, scaled by a penalty parameter $\lambda_c^{(\kappa)}$ for smoothness. Then, $\log(\lambda_c^{(\kappa)}) \sim \mbox{Normal}(\upsilon^{(\kappa)},\sigma_{2}^{2(\kappa)})$. Unlike the splines for $\nu_{i,j,c,t}$, the prior expectations are zero, as opposed to regional or super-regional. This is because we have no prior belief that residual deviation from UN estimates in the sampling of urban respondents should be regionally structured. Employing thin-plate splines here allows $g_c (t)$ to capture non-linear deviations from $P_{c,t}$ over time, but only when there is ample evidence in the data for a given country.

\subsection{Robustness to outliers}\label{sec:outliers}
In addition to as the main data-specific modelling challenges highlighted in Section \ref{sec:method}, the database contains some recorded values which truly defy the observed trend in their country. These values often can't be explained by normal survey variability alone, and can have an undue influence on the estimated trend if treated like ordinary observations. While the Beta-Binomial conditional models we employ are already more robust to outliers than equivalent Binomial models, severe outliers can still cause issues, including causing the estimated trend to deviate substantially from other surveys to be closer to the outlier, or the over-estimation of survey variability.

To address this problem, we model each observation as arising from a mixture distribution, which combines the Beta-Binomial conditional model with a discrete Uniform distribution. The extent to which the model is either Beta-Binomial or Uniform is controlled by the mixing parameter $\rho$: as $\rho$ approaches 0, the mixture becomes Beta-Binomial and vice-versa:
\begin{eqnarray}
p(v_i\mid v_{-i}, \bm{\alpha}, \bm{\beta}) &=& \rho \binom{n_i}{v_i} \frac{B(v_i+\alpha_i ,n_i-v_i+\beta_i)}{B(\alpha_i , \beta_i)} + (1-\rho)\frac{1}{n_i}. 
\end{eqnarray}
This approach effectively allows the model to decide, given sufficient evidence in the data, whether or not a survey observation could plausibly have arisen from the same model as other nearby (in time) surveys for that country and area. The degree of evidence required can be controlled through the prior distribution specified for each $\rho$. For example, a strong prior distribution with most of the probability mass close to 0 for each $\rho$ corresponds to a strong belief that each survey value is very unlikely to be an outlier. 

For this application, we introduce one $\rho$ for each unique survey. This means that if a survey has an urban, rural, and an overall value, a single $\rho$ controls the extent of mixing for all three. The reason for this is that if, for example, the model indicates an urban value is a very severe outlier, we would prefer to also reduce the effect of the corresponding rural value on estimated trends and uncertainty. Including this layer in the model means that estimated trends are considerably more robust to outliers, as we will highlight in Section \ref{sec:check}. Additionally, predictions for $\rho$ are useful as an indicator to efficiently flag surveys that may warrant further investigation.

\subsection{Prior distributions and implementation}\label{sec:priors}
For all hyper-parameters $\upsilon$ which are the mean of a Normal distribution (e.g. $\upsilon_{i,j}^{(\beta)}$), we specified non-informative Normal($0,10^2$) prior distributions. For all hyper-parameters $\sigma$ which are the standard deviation of a Normal distribution (e.g. $\sigma_{0,i,j}^{(\beta)}$), we specified non-informative positive-truncated Normal($0,10^2$) prior distributions.

All code was written and executed using R (\cite{R}) and the model was implemented using NIMBLE \citep{nimble}, a facility for highly flexible implementation of Markov Chain Monte Carlo (MCMC) models. For this application, we needed to add the Beta-Binomial distribution to NIMBLE, which was straightforward using only a few lines of R code. Four MCMC chains were run for 80,000 iterations from different randomly generated initial values and with different random number generator seeds. The first 40,000 samples were discarded as burn-in and, to limit system memory usage, the remaining samples were thinned by 10. Convergence of the MCMC chains is discussed in Appendix \ref{app:convergence}. The model was applied to a subset of the data consisting of 1084 surveys and predictions were made for all countries with at least one survey (after selection). Survey selection criteria are discussed in Appendix \ref{app:selection}. Associated NIMBLE model code is included as supplementary material and data is available on request.

\section{Model Checking}
\label{sec:valid}
The task of assessing the validity of the statistical model is divided into two parts: basic procedures to check there are no systematic issues with reproducing the observed data and a forecasting experiment to evaluate the ability of the model to predict future fuel usage values. 

\subsection{Posterior predictive checking}\label{sec:check}
Given the Bayesian implementation of the model, assessing the fit to both in-sample and out-of-sample data is based on posterior predictive model checking \citep{Gelman2013}. For in-sample data, this involves using samples from the joint posterior distribution of parameters and random effects (which are already available from MCMC) to simulate $v_i$ from the conditional Multinomial distribution. This results in samples from the posterior predictive distribution for replicates $\bm{\tilde{\bm{x}}}|\bm{x}$ of the observed fuel proportions $\bm{x}$. The statistical properties of these replicates can then be compared to properties of the corresponding observations. For brevity, we present predictive checking for solid fuel use in this subsection and for all of the other fuel types in Appendix \ref{app:further_check}.
\begin{figure}[h!]
\floatbox[{\capbeside\thisfloatsetup{capbesideposition={right,center},capbesidewidth=0.3\linewidth}}]{figure}[1.36\FBwidth]
{\caption{Scatter plots comparing posterior means of solid fuel usage replicates $\tilde{x}_{1,j,c,t}$ to their corresponding observed values.}
\label{fig:wfit}}
{\includegraphics[width=1\linewidth]{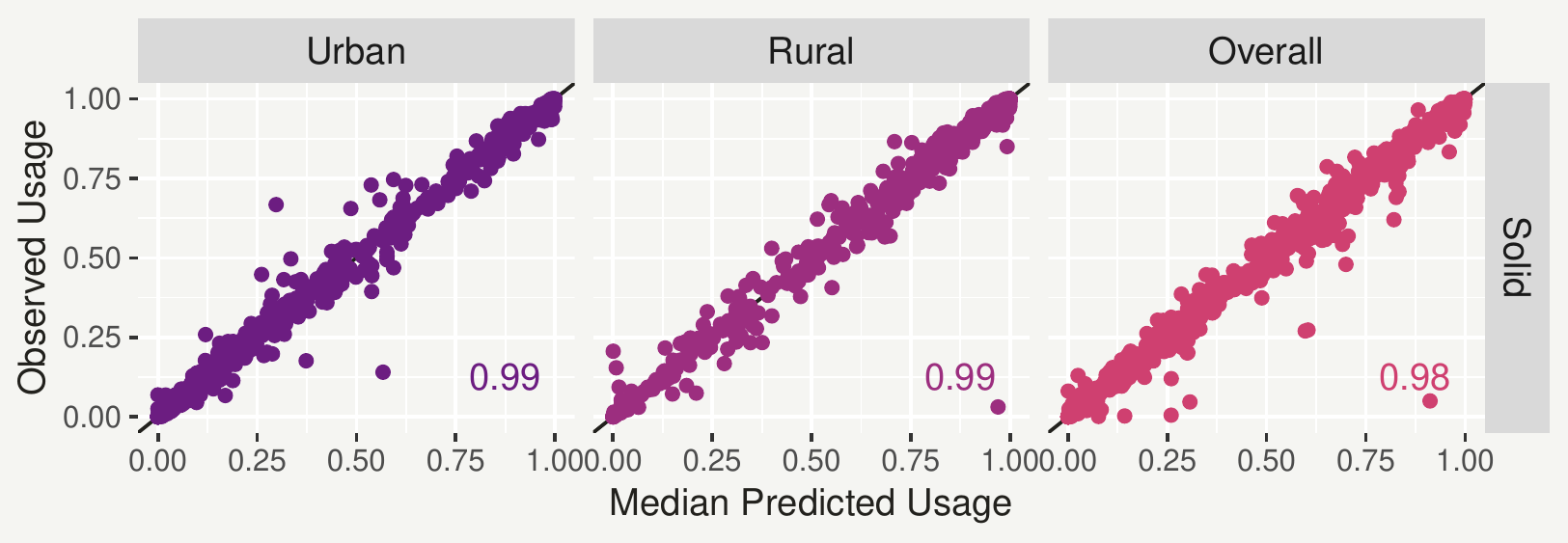}}
\end{figure}

In the first instance, scatter plots comparing the posterior means of the replicates with the observed values can give an indication of any systematic issues. These are shown for solid fuels in Figure \ref{fig:wfit} and, for the most part, there are no obvious systematic problems. Also shown are coverage values: the proportion of observed solid fuel use values which lie within the 95\% posterior predictive intervals, computed from the corresponding replicates. A coverage substantially lower than 95\% would mean a high proportion of observed values are extreme values with respect to the posterior predictive model, implying a poor fit. In this case, the coverage values for the 95\% credible intervals were higher than 95\% for all fuels and areas. Taken together, these two checks indicate that the model captures the observed data well.
\begin{figure}[h!]
\includegraphics[width=\linewidth]{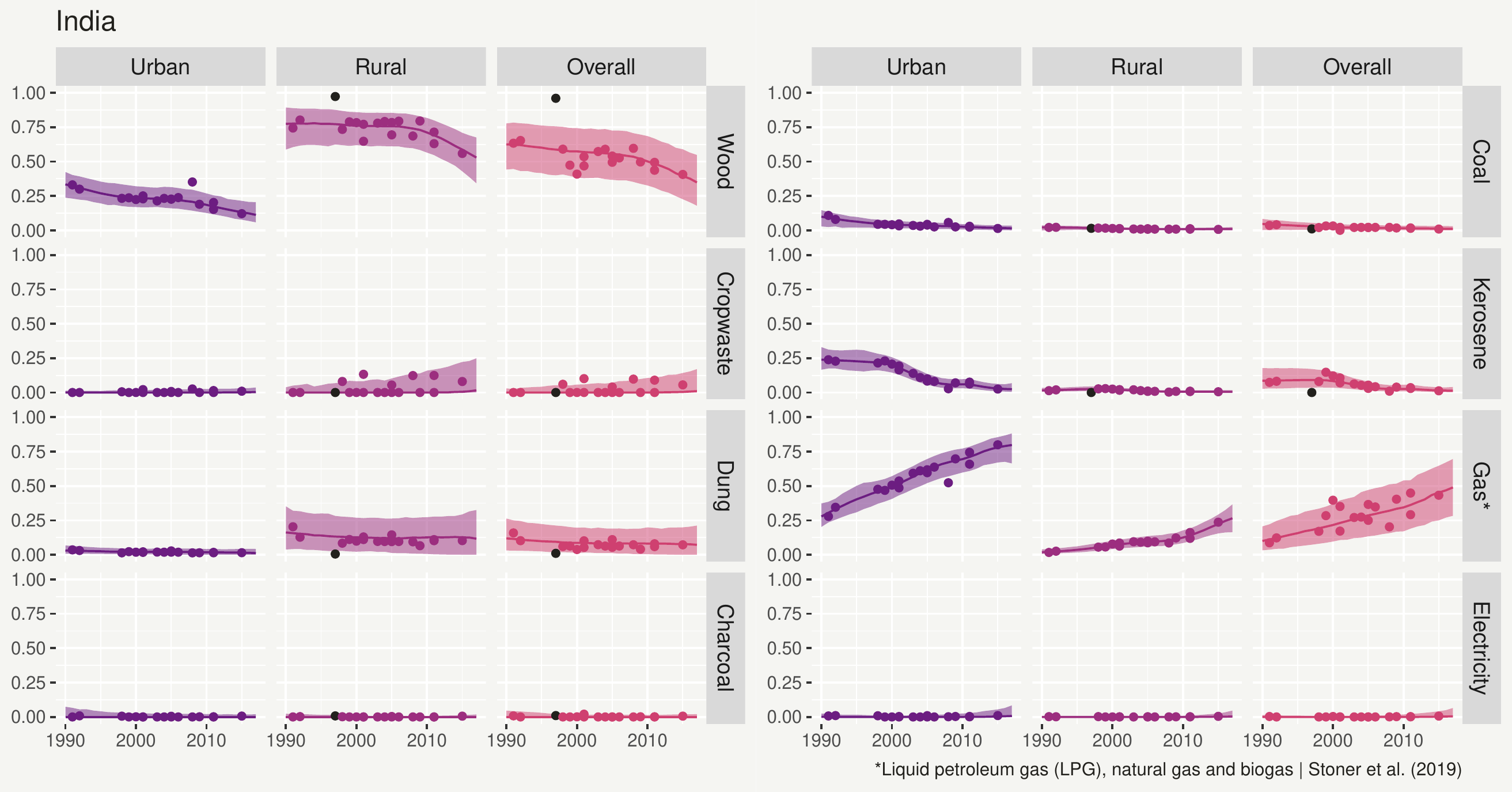}
\caption{Predicted fuel usage trends (median and 95\% prediction intervals) for India. Coloured points are survey observations and black points are removed surveys. For each fuel, the left, central and right plots show urban, rural and overall usage, respectively.}
\label{fig:india}
\end{figure}

Another way of checking the model is to compare predicted trends to survey observations on an individual country basis. Figure \ref{fig:india} shows the median predicted proportion using each fuel in each segment (urban, rural and overall) of India, with associated 95\% posterior predictive intervals. Here it can be seen that the predicted trends follow the observed trends well, with prediction intervals that envelop a reasonable number of surveys. Moreover, by examining the tightness of the prediction intervals with respect to the variance of the observations, we can see that the high coverage values obtained for the replicate prediction intervals are not simply caused by excessively high model uncertainty.  
\begin{figure}[h!]
\includegraphics[width=\linewidth]{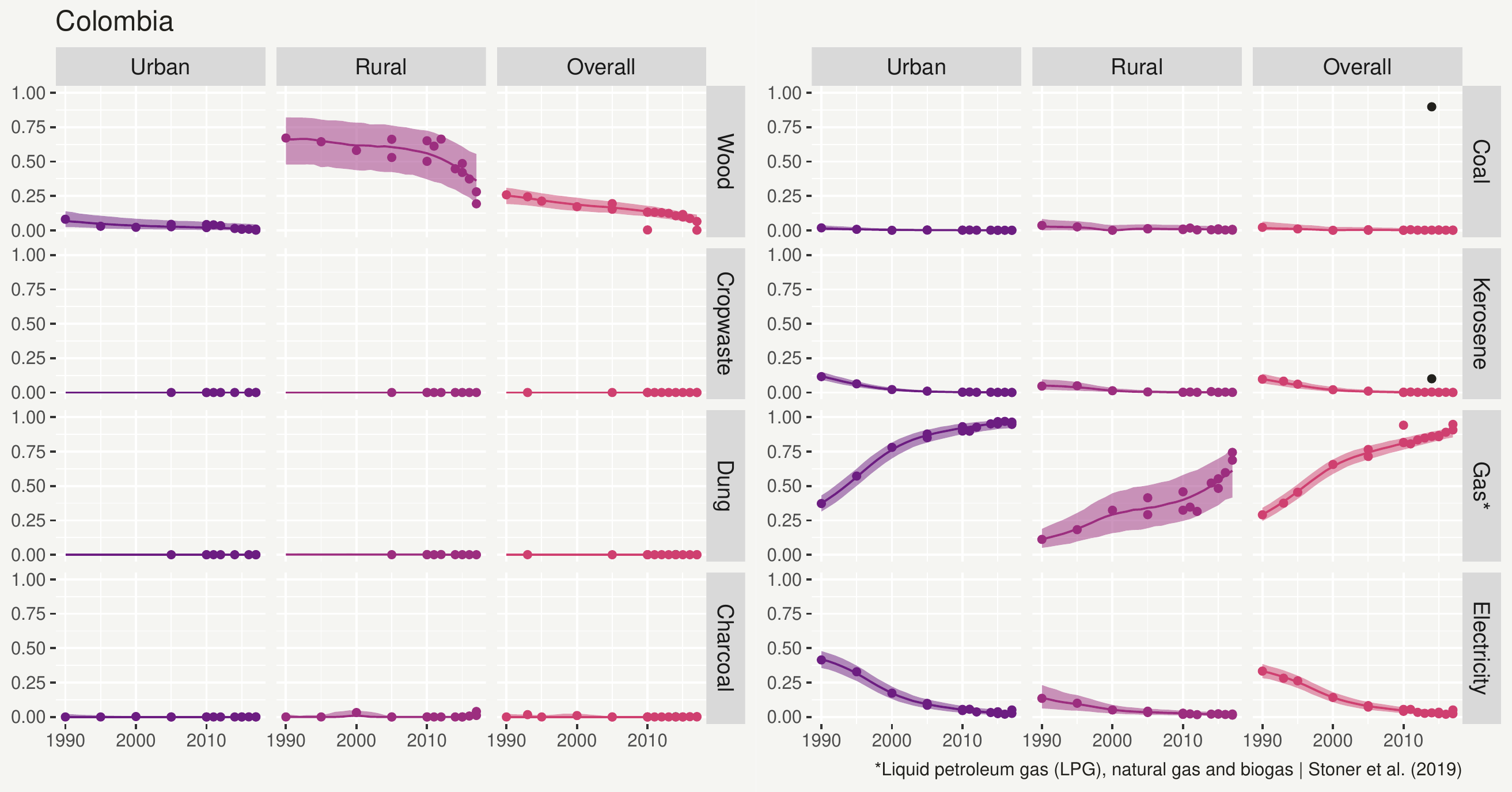}
\caption{Predicted fuel usage trends (median and 95\% prediction intervals) for Colombia.}
\label{fig:colombia}
\end{figure}

A similar plot is shown for Colombia in Figure \ref{fig:colombia}. Looking at the use of gas in 2010, we can see there is one survey with an unusually high overall value. Through the $\rho$ corresponding to this survey, the model suggests that this value is likely an outlier, such that the estimated trend and variability are not adversely affected. This illustrates the effectiveness of incorporating mixture distributions (as described in Section \ref{sec:outliers}) in making the model more robust to outliers.

Note that to check whether the model reproduces the observed data well, the overall predictions in Figures \ref{fig:india} and \ref{fig:colombia} incorporate the model's prediction of any systematic deviation ($g_c(t)$) from the UN estimates of urban and rural proportions, in the sampling of urban and rural respondents. If desired, predictions of overall fuel usage can instead be based solely on the UN estimates of urban and rural proportions (rather than based on the proportions in the surveys). This is achieved by removing $g_{c}(t)$ from \eqref{eq:urban_proportion} during simulation. 

Predicted fuel usage plots which include survey sampling variability (as in Figures \ref{fig:india} and \ref{fig:colombia}) are included as supplementary material for the 8 most populous countries (as of late 2019, excluding the US and Russia).

\begin{figure}[h!]
\floatbox[{\capbeside\thisfloatsetup{capbesideposition={right,center},capbesidewidth=0.3\linewidth}}]{figure}[1.36\FBwidth]
{\caption{Predicted mean survey urban proportions for Kenya (left) and Malawi (right), compared to observed survey urban proportions and associated U.N. urban population estimates.}
\label{fig:urban_proportion}}
{\includegraphics[width=1\linewidth]{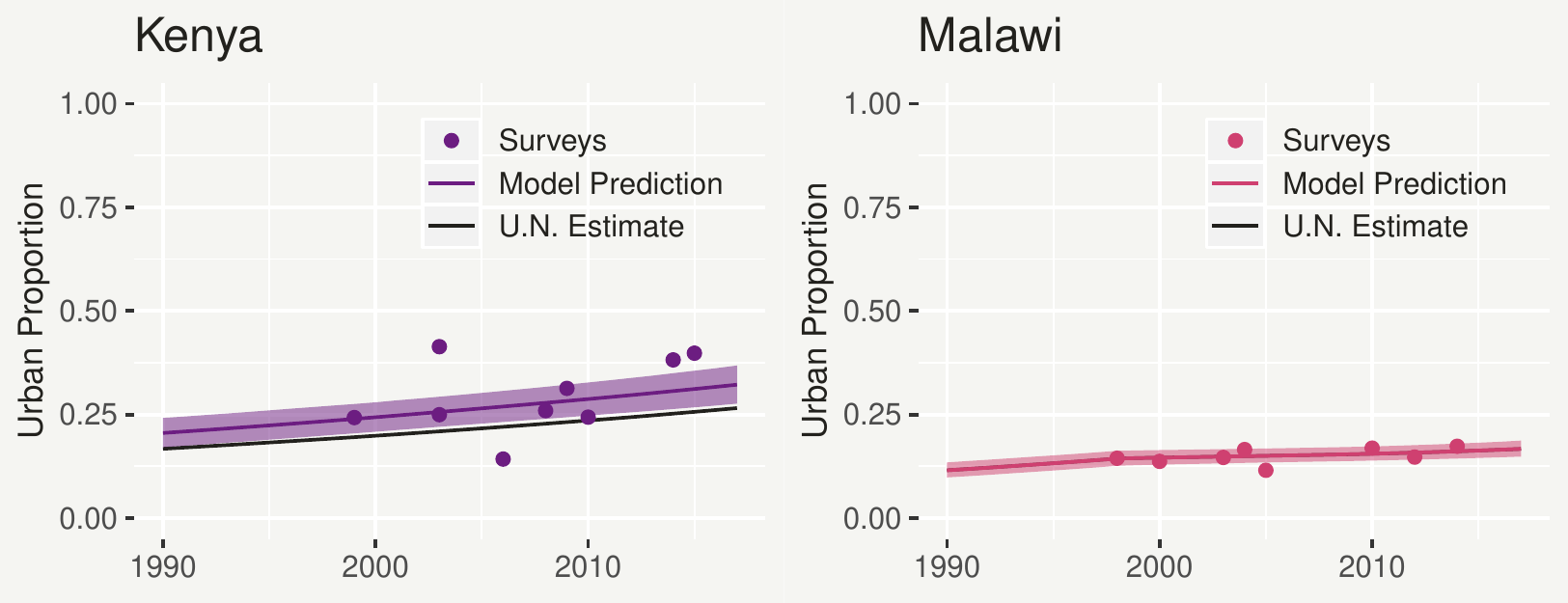}}
\end{figure}

We can also inspect the model's ability to capture structured between-country and temporal variability in the proportions of urban and rural respondents in the survey samples: Figure \ref{fig:urban_proportion} shows the proportion of (unweighted) respondents recorded as urban in the fuel surveys for Kenya (left) and Malawi (right) compared to UN estimates and predicted values from the model. The plot for Kenya shows evidence that the proportion of urban respondents in the surveys is, on average, higher than the UN estimates ($g_c(t)>0$). The plot for Malawi, meanwhile, shows limited evidence of any systematic deviation ($g_c(t)\approx 0$). In both of these cases, the spline incorporated in \eqref{eq:urban_proportion} appears to capture any remaining structured variability (or the lack thereof) well, enabling reliable prediction of urban and rural trends where surveys only provide values for the overall population.

\subsection{Forecasting experiment} \label{sec:pred}
The model's ability to predict (forecast) fuel usage beyond the range of the data can be assessed using out-of-sample predictive testing. This is important to validate the model's use for predicting present-day fuel use, as there is a lag in data collection of 1-2 years, and for projecting estimated trends into the future, to provide a baseline against which the effects of interventions can be compared. To emulate a hypothetical forecasting scenario, the model was fitted only to surveys up to and including year 2012, therefore excluding 5 years (approximately 22\% of the data). We then used the model to predict 5 years into the future and produce predictive distributions for the out-of-sample surveys. As it is not our primary interest to forecast how any systematic trends in the sampling of urban and rural respondents will progress in the future, we focus on checking the out-of-sample prediction of urban and rural surveys.

\begin{figure}[h!]
\includegraphics[width=\linewidth]{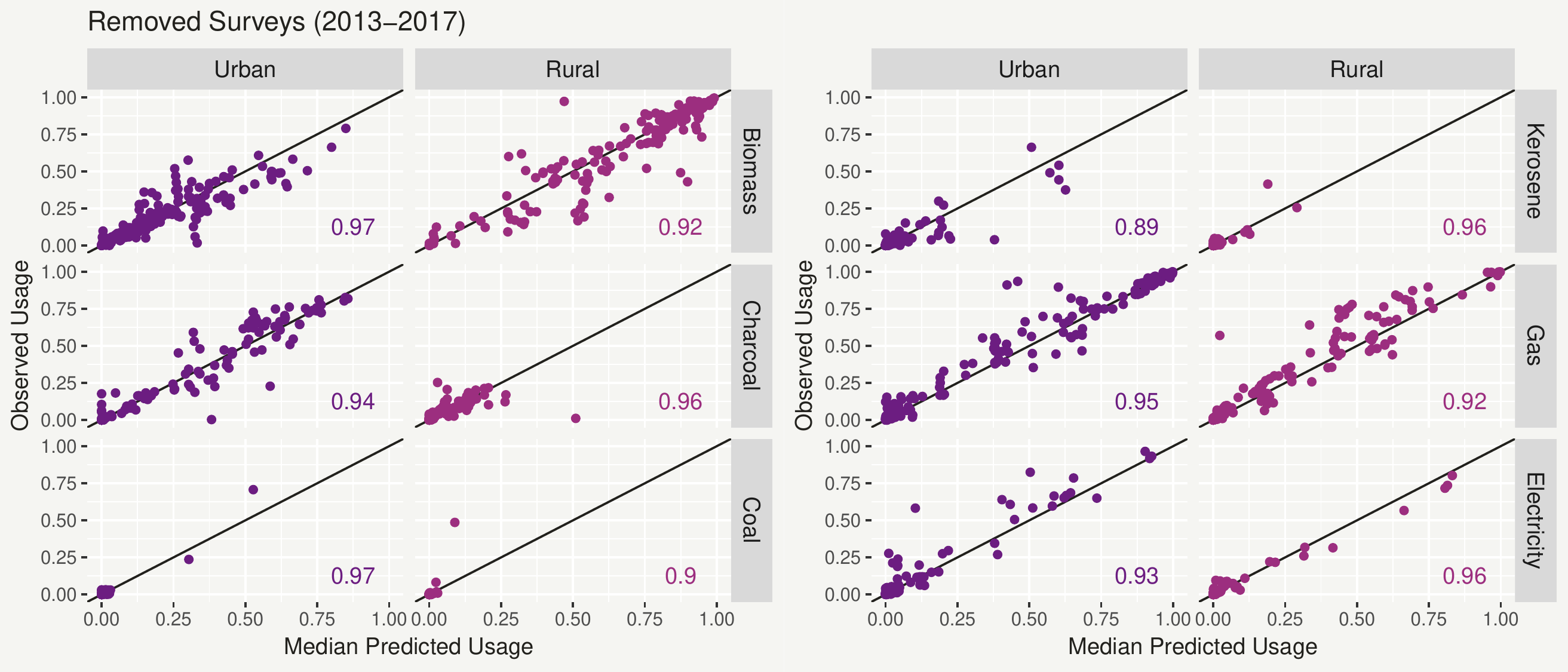}
\caption{Scatter plots of mean predicted fuel usage values from 2013 onwards, versus their observed values, from the model which was only supplied data from 2012 or earlier.}
\label{fig:validation}
\end{figure}

Figure \ref{fig:validation} shows scatter plots comparing the out-of-sample survey values to the mean predicted values from the model. While there are some values which are not captured well (some potentially due to errors in data entry), generally the model does not seem to systematically over or under-predict. Notably, the coverage values tend to be quite high, indicating that the model produces reliable uncertainty estimates when predicting into the future. 

\begin{figure}[h!]
\includegraphics[width=\linewidth]{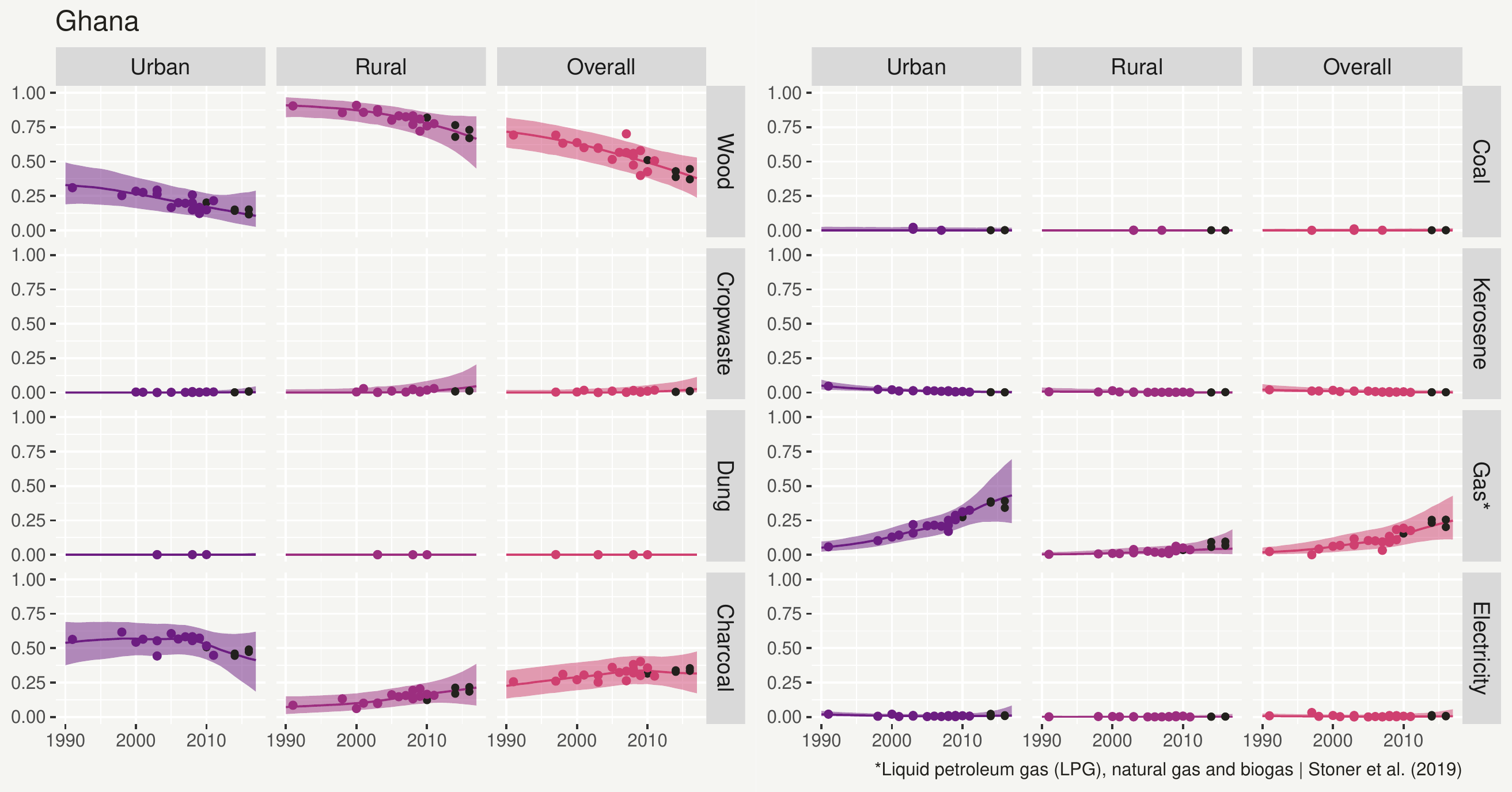}
\caption{Predicted fuel usage trends (median and 95\% prediction intervals) for Ghana, from the model where surveys from 2013 onwards were excluded. The black points from 2013 onwards show excluded surveys.}
\label{fig:ghana}
\end{figure}

To guard against high coverage values through unreasonably uncertain prediction intervals,  we can assess the model's performance when forecasting by examining predictive plots for individual countries. Figure \ref{fig:ghana} shows predictive fuel usage plots for Ghana, from the model where surveys from 2013 onwards  are excluded. Here, the removed surveys are generally well within the 95\% predictive intervals, which grow reasonably larger for predictions further into the future, but are not so wide that they are impractical. 

\section{ Discussion}
\label{sec:conclusion}
Currently, the  health burdens associated with  exposure to air pollution from the use of polluting fuels for cooking are assessed based on groupings of fuel types (i.e. solid fuels or polluting fuels). However, this fails to take into account changes in the use of specific fuel types that may affect the impacts on health. For example, the results of the analyses performed here suggest  that over the last few decades a substantial proportion of urban households in Sub-Saharan Africa have switched from raw biomass fuels (i.e. wood, cropwaste and dung) to charcoal, which has very different emissions characteristics. To expand the knowledge base about the impacts of air pollution on health, burden of disease calculations should instead be based on the use of specific fuels, but until now country-specific  estimates of specific fuel usage have been unavailable.

To address this, we have developed and implemented a multivariate hierarchical model for specific fuel types which aims to: (i)  estimate trends and associated measures of uncertainty,  for specific fuels, for every country, and separately for urban and rural areas within a coherent modelling framework; (ii) provide meaningful estimates in countries where there is limited data; (iii) forecast fuel usage up to present day and into the future. 

Based on Generalized Dirichlet Multinomial distributions, the Global Household Energy Model (GHEM) automatically constrains the proportions of populations using each of eight key fuel types ensuring that their sum does not exceed one. Set within a Bayesian modelling framework, parametric and predictive uncertainty is  quantified (e.g. by 95\% prediction intervals) and  verified using within-sample posterior predictive checking (see Section \ref{sec:check}). Where  data availability is limited  within a country, the model  is able to `borrow' information from neighbouring countries using  nested country, regional and super-regional random effects, reducing predictive uncertainty. The model can forecast a number of years beyond the extent of the data, with  assessment of forecasted values performed  using   an out-of-sample predictive experiment (see Section \ref{sec:valid}). This allows  present-day fuel use to be evaluated, as data collection lags behind by 1-2 years. In addition, fuel use predictions for future years provide a baseline representation of what might be expected  in the absence of intervention, to  which future surveys conducted post-interventions can be compared.

In achieving these aims, the model overcomes a number of challenges associated with using these survey data: (a) inconsistency in survey design and collection, together with missing values, which can lead to highly unstable time series for some individual fuels in some countries; (b) the total number respondents is unavailable for around half of surveys; (c) for many surveys, fuel use values are not available separately for urban and rural areas. 

To address (a), we adopted a tiered approach (Section \ref{sec:tiers}) where we first modelled combined fuel use (e.g. solid fuels), which is   progressively disaggregated into the component fuels. This ensures that excess variability and uncertainty among `confused' fuels does not propagate into those unaffected, and that predictions for the aggregate quantities are stable. In order to address the problem where the total number of respondents is unknown, (b), we approximate a GDM model for the number of respondents,  transforming the proportions using each fuel into counts from an artificial sample size (Section \ref{sec:GDM}). We  illustrate that this results in approximately the same inference for population-wide fuel usage as modelling the (unavailable) number of respondents in their original count form, through a simulation experiment (presented in Appendix \ref{app:simulation}). We addressed the unavailability of information on separate urban and rural fuel use  for all surveys, (c),  by including a layer in the model which links the urban, rural and overall fuel use values for each survey. Structured between-country and temporal variability in the proportion of urban respondents was then accounted for by combining UN estimates with  smooth functions of time for each country. Finally, in addition to addressing these data-specific challenges, mixture distributions were employed to make the model more robust to potential outliers.

To date, the model has been adopted by the WHO to produce estimates of the proportion of people in each country who rely on polluting fuels as their primary fuel and technology for cooking and has played a central role in  monitoring  SDG 7.1.2 \citep{trackingreport}. It has also played an important role in identifying data that appear to be out-of-line with general country-level patterns for further investigation. Ultimately, the  proposed  modelling approach provides policy-makers with decision-quality information and   enables a ground-breaking  re-assessment of the health impacts of  cooking  with polluting fuels and technologies.

\section*{Acknowledgments}
This work was supported by a Natural Environment Research Council GW4+ Doctoral Training Partnership studentship [NE/L002434/1] and the World Health Organization contract APW 201790695.

\begin{appendices}
\section{Simulation Experiment}\label{app:simulation} 
To illustrate the validity of our approximation for modelling the proportions using each fuel type $\bm{x}=\bm{y}/n$, we present a simulation experiment using the 598 observed survey samples sizes $n$. The majority are in the range 1000-100000, with a mode of around 10000. At these large values, the contribution of the Multinomial variance to the total variance of $\bm{x}$ would be small. 

For each available $n_i$ ($i=1,\dots,598$), we simulate a vector of survey responses $\bm{y}_i=\{y_{i,1},y_{i,2},y_{i,3},y_{i,4}\}$ from a GDM model. Here, each country has a different (time constant) marginal mean vector $\bm{\mu}_c$ and variance parameters $\bm{\phi}_c$ (preserving the original associations between the countries and observed $n_i$ in the data, and ignoring countries with no observed $n_i$) . Note that some countries will only have one $\bm{y}_i$ and others will have several (each with its own unique $n_i$). We simulate all of the $\bm{\mu}_c$ from a Dirichlet($\bm{1}$) distribution, and all of the $\bm{\phi}_c$ independently from a Gamma($4,0.1$) distribution (inducing a moderately high degree of over-dispersion, compared to the Multinomial):
\begin{align}
\bm{y}_{i} \sim  \mbox{GDM}(\bm{\mu}_c,\bm{\phi}_c,n_i); \qquad
\bm{\mu}_c \sim  \mbox{Dirichlet}(\bm{1}); \qquad
\bm{\phi}_c \sim  \mbox{Gamma}(4,0.1).
\end{align}

In the baseline scenario, to which we will compare our approximate method, we have observations for all of the $n_i$ and all of the $\bm{y}_i$. This allows us to implement the above model directly, which we do in a Bayesian setting using a Dirichlet($\bm{1}$) prior for each $\bm{\mu}_c$ and a non-informative Exponential(0.001) prior for each $\phi_c$.

In the second scenario, we don't know any of the $n_i$ or the $\bm{y}_i$, but we do have observations for $\bm{x}_i=\bm{y}_i/n_i$. In this scenario, we can apply our approximate method (from Section \ref{sec:GDM}), where we fit the GDM to constructed counts $\bm{v}_i= \lfloor N \bm{x}_i \rfloor$. We proceed to apply this method whilst varying $N$ over a range of values (10, 20, 30, 50, 100, 300, 1000, 3000, 10000, 30000, 100000, 300000, 1000000), so that we can investigate the impact of this choice on parameter inference.
\begin{figure}[h!]
\floatbox[{\capbeside\thisfloatsetup{capbesideposition={right,center},capbesidewidth=0.55\linewidth}}]{figure}[0.85\FBwidth]
{\caption{The top panel shows the median, interquartile range (dark) and 95\% interval (light) of the mean squared differences between the posterior samples of the marginal mean proportions $\mu_{1,c},\dots,\mu_{4,c}$ and their corresponding true values, from the approximate model with varying $N$. Similarly, the bottom plot shows the median, interquartile range and 95\% interval of the posterior standard deviations of $\mu_{1,c},\dots,\mu_{4,c}$. The dashed lines represent these results from the baseline model.} \label{fig:experiment_plots}}
{\includegraphics[width=1\linewidth]{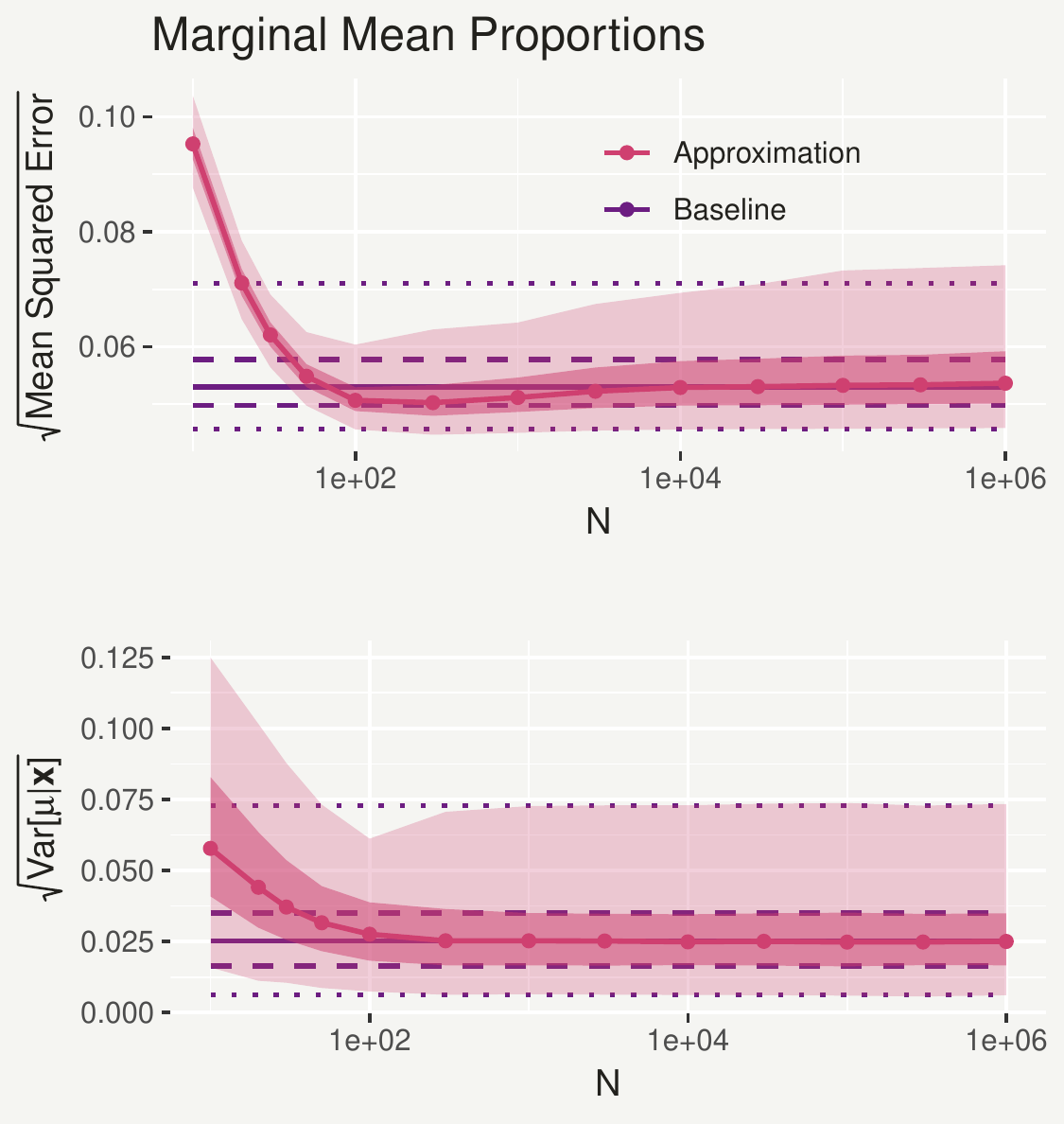}}
\end{figure}

Recall that in our application we are primarily interested in correct inference for the marginal mean proportions $\bm{\mu}_c$ (the population-wide fuel use in each country), and we claimed that a sufficiently large choice of $N$ yields a parameter inference approximately the same as if we had modelled the $\bm{y}_i$ directly, along with the sample sizes $n_i$. To assess this, we begin by examining the models' accuracy when predicting the true marginal mean proportions $\bm{\mu}_c$. For each posterior sample, we can compute the mean squared error between the predicted values of $\bm{\mu}_c$ and the true values. The top panel of Figure \ref{fig:experiment_plots} shows the median of this statistic, for varying $N$, as well as the inter-quartile range (dark), and 95\% prediction interval (light). Compared to the same statistics for the baseline model, shown as horizontal lines, we can see that the distribution of mean squared errors for the approximate method does indeed converge to the baseline model as $N$ increases, from about $N=10000$ onwards.

\begin{figure}[h!]
\includegraphics[width=\linewidth]{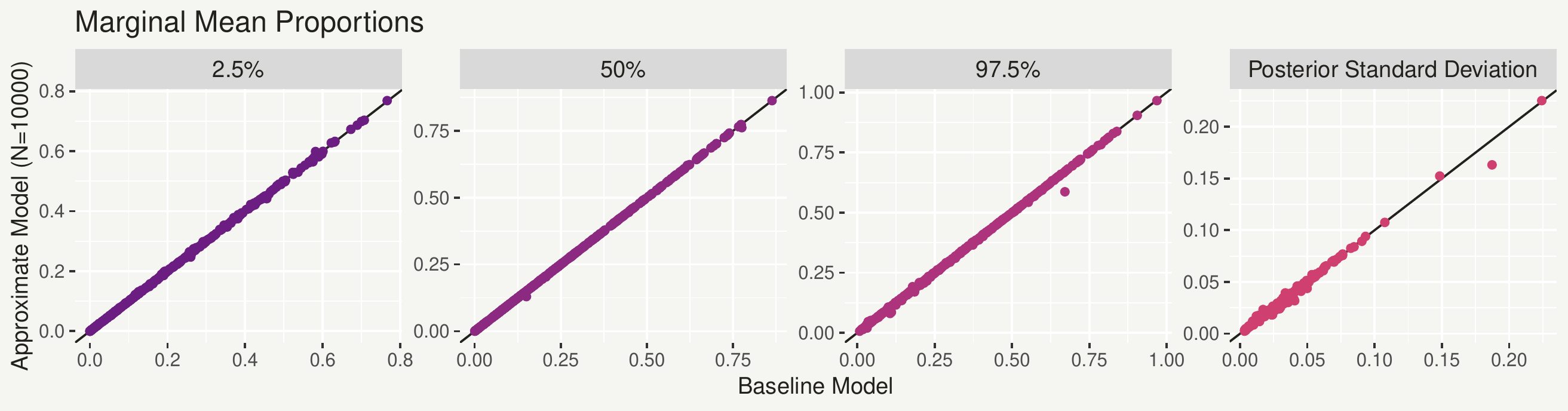}
\caption{Scatter plots comparing the posterior 2.5\%, 50\%, and 97.5\% posterior quantiles, and posterior standard deviations for the marginal mean proportions $\mu_{1,c},\dots,\mu_{4,c}$, from the approximate model with $N=10000$, to the baseline model.}
\label{fig:mu_quantiles}
\end{figure}
We can also examine how the approximate method quantifies uncertainty in $\bm{\mu}_c$. For each individual $\mu_{1,c},\dots,\mu_{4,c}$, we compute the standard deviation of the posterior samples. The median of these posterior samples are then shown for each $N$ in the bottom panel of Figure \ref{fig:experiment_plots}, once again alongside the inter-quartile range and 95\% interval. The distribution of posterior standard deviations for the approximate method also converges to the baseline model, but does so for a much lower $N$ (between 100 and 1000) than the mean squared error.

Finally, if we choose a single value of $N$, we can compare more closely the approximate method to the baseline model when estimating $\bm{\mu}_c$. Figure \ref{fig:mu_quantiles} compares the 2.5\%, 50\%, and 97.5\% posterior quantiles for the $\mu_{1,c},\dots,\mu_{4,c}$ from the approximate model with $N=10000$, to the quantiles from the baseline model. The quantiles are virtually identical, suggesting that for this simulated data the same inference for $\bm{\mu}_c$ would be achieved either by modelling the true counts $\bm{y}_i$ directly or by modelling the constructed counts $\bm{v}_i= \lfloor 10000 * \bm{x}_i \rfloor$.

\section{Convergence of MCMC Chains}\label{app:convergence}
One way to assess the convergence of MCMC chains is to compute the Potential Scale Reduction Factor (PSRF) for a number of key parameters. This compares the variance between the MCMC chains to the variance within the chains \citep{convergence}. A PSRF of 1 is obtained when the two variances are the same, so starting the chains from different initial values and obtaining a PSRF close to 1 (typically taken to be less than 1.05) gives a good indication that the chains have converged to the parameter's posterior distribution. 
\begin{figure}[h!]
\floatbox[{\capbeside\thisfloatsetup{capbesideposition={right,center},capbesidewidth=0.4\linewidth}}]{figure}[1.15\FBwidth]
{\caption{Histograms of the Potential Scale Reduction Factor (PSRF) for the relative means $\nu_{i,j,c,t}$ and variance parameters $\phi_{i,j,c}$.}
\label{fig:psrf}}
{\includegraphics[width=1\linewidth]{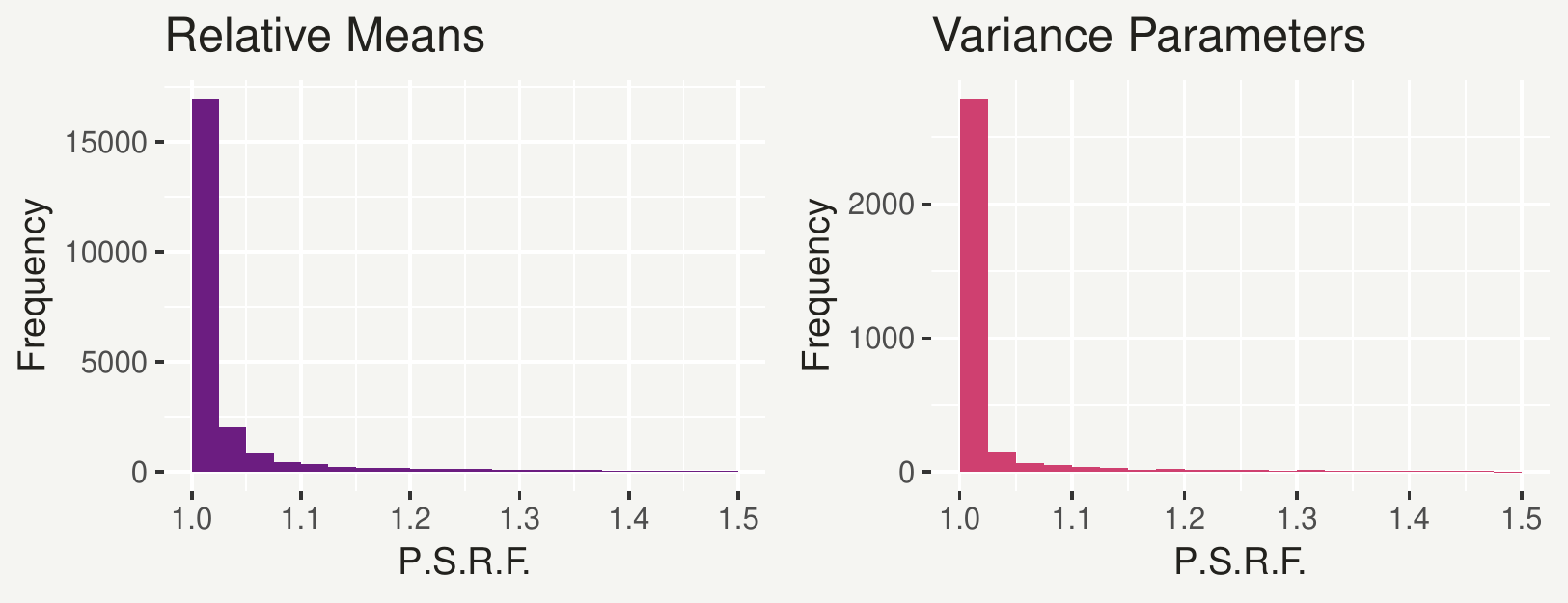}}
\end{figure}

We computed the PSRF for the (26016) relative means $\nu_{i,j,c,t}$ corresponding to the survey observations and the (3576) variance parameters $\phi_{i,j,c}$. Figure \ref{fig:psrf} presents these respectively in frequency histograms. For both sets of parameters, the overwhelming majority of the values lie in the closest bin to 1, suggesting that the model has converged.

\section{Further Model Checking}\label{app:further_check}
As discussed in Section \ref{sec:check}, it is important to verify that the model is able to reproduce the observed data well. We do this by comparing replicates (predictions) of the observed data to the actual observations. In Section \ref{sec:check} we checked the replicates of solid fuel use and here we check the remaining fuels.
\begin{figure}[h!]
\floatbox[{\capbeside\thisfloatsetup{capbesideposition={right,center},capbesidewidth=0.3\linewidth}}]{figure}[1.36\FBwidth]
{\caption{Scatter plots comparing the posterior means of kerosene, gas and electricity use replicates to their corresponding observed values.}
\label{fig:top_fit}}
{\includegraphics[width=1\linewidth]{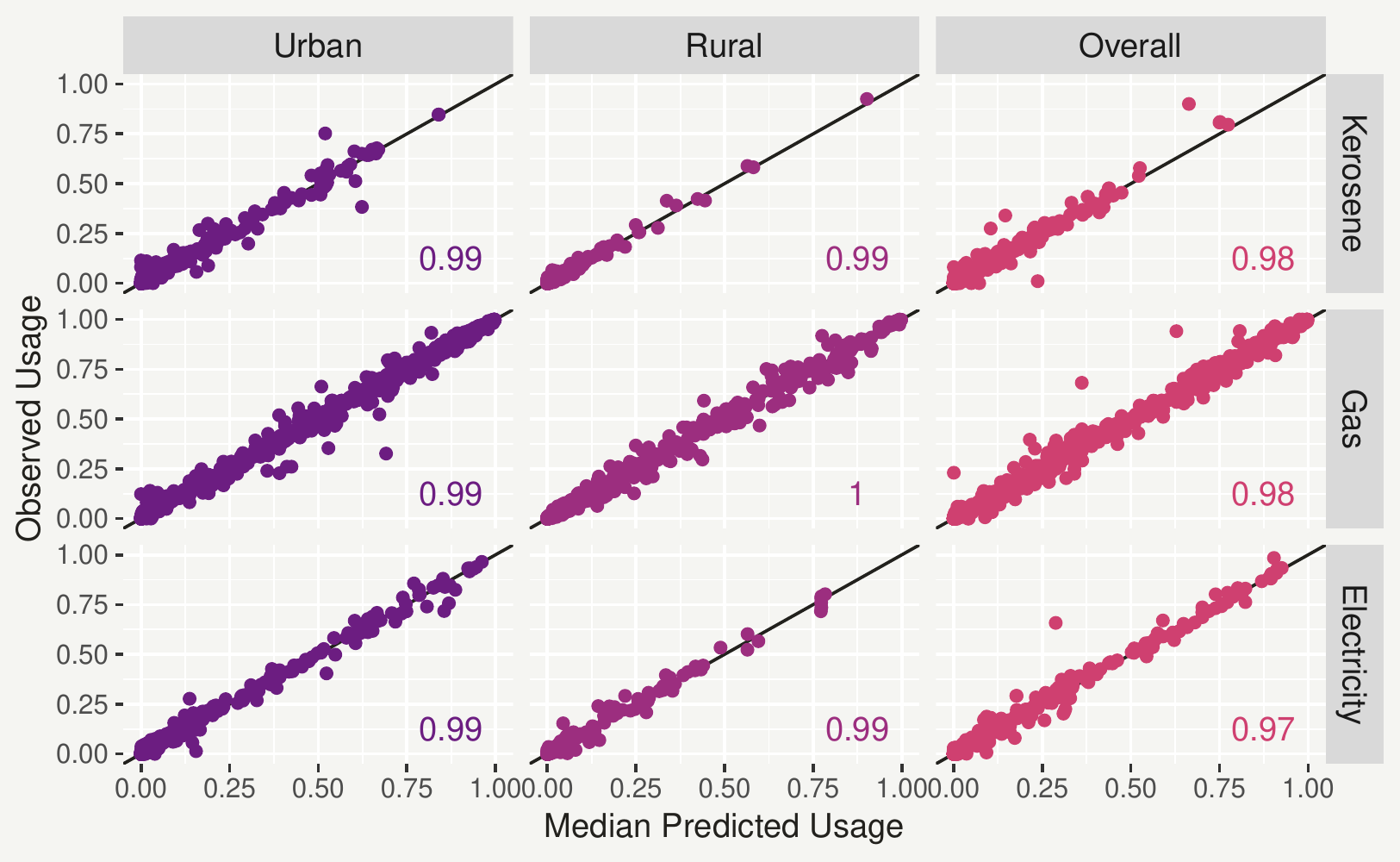}}
\end{figure}

\begin{figure}[h!]
\floatbox[{\capbeside\thisfloatsetup{capbesideposition={right,center},capbesidewidth=0.3\linewidth}}]{figure}[1.36\FBwidth]
{\caption{Scatter plots comparing the posterior means of biomass, charcoal and coal use replicates to their corresponding observed values.}
\label{fig:mid_fit}}
{\includegraphics[width=1\linewidth]{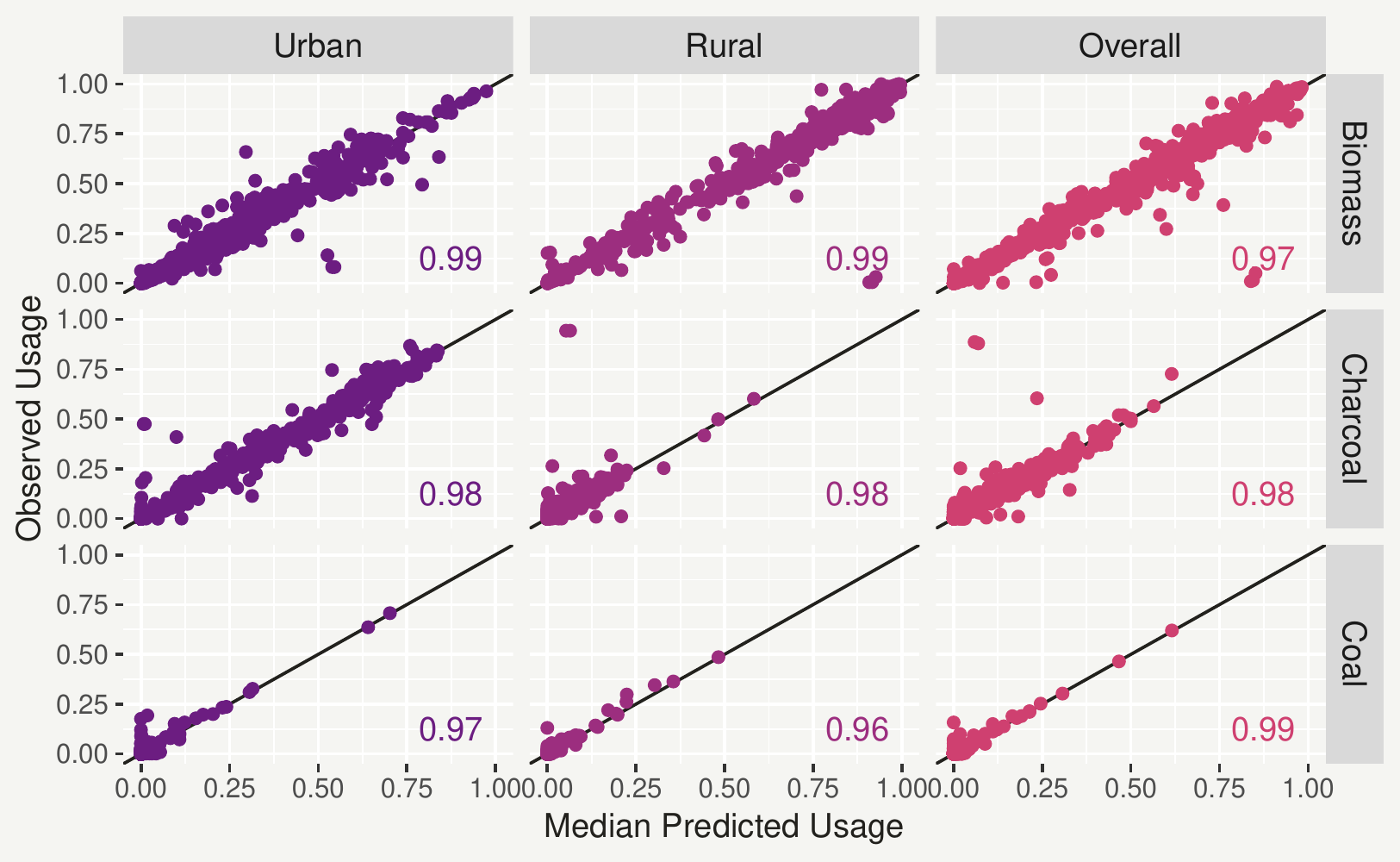}}
\end{figure}

\begin{figure}[h!]
\floatbox[{\capbeside\thisfloatsetup{capbesideposition={right,center},capbesidewidth=0.3\linewidth}}]{figure}[1.36\FBwidth]
{\caption{Scatter plots comparing the posterior means of wood, cropwaste and dung use replicates to their corresponding observed values.}
\label{fig:lower_fit}}
{\includegraphics[width=1\linewidth]{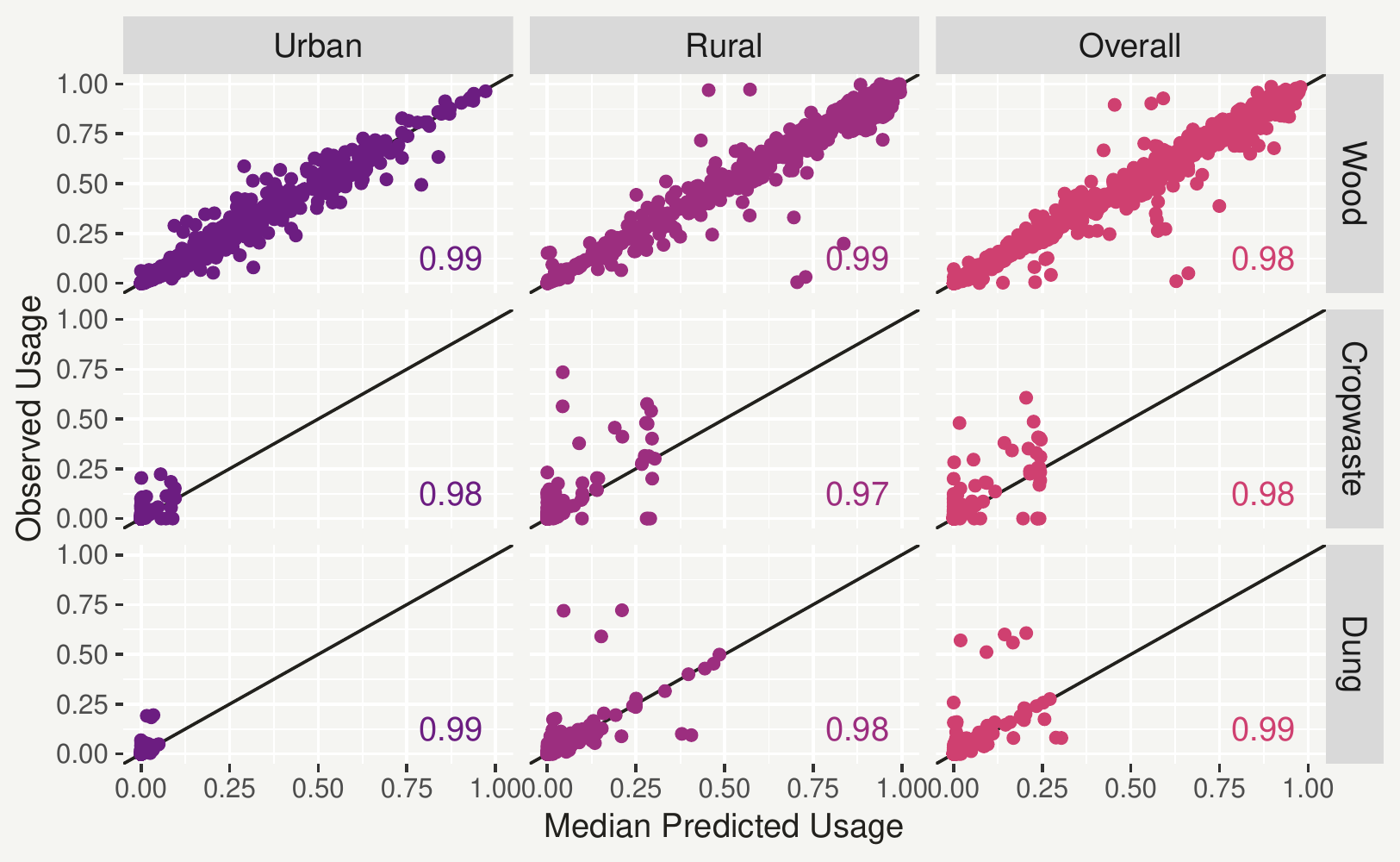}}
\end{figure}

Figure \ref{fig:top_fit} shows scatter plots comparing the mean predicted replicates for the three other main top-tier fuel types, kerosene, gas and electricity, to their corresponding observed values. Similarly, Figure \ref{fig:mid_fit} shows the same plots for the three mid-tier fuel types, biomass, charcoal and coal, and Figure \ref{fig:lower_fit} shows the three lower-tier fuel types, wood, cropwaste and dung. In general the points are scattered about the diagonal line fairly evenly, indicating a good model fit for the different fuels. Notably, however, the fit of the model is more precise for fuel types in the upper tiers (e.g. electricity) than those in the lower tier (e.g. dung). This makes sense, as these fuels are less likely to be affected by the issues described in Sections \ref{sec:method} and \ref{sec:conditional}, such as the combination of certain fuel types, where some of the observed values are likely to be erroneous and difficult for the model to capture well. Regardless, the coverage of the 95\% intervals is very high for all fuels. 

\section{Survey Selection}\label{app:selection}
The model was applied to a selection of the WHO Household Energy Database. Surveys were excluded from the analyses if they: 
\begin{itemize}
\item only reported the usage of `solid fuels' as a group, rather than the usage of at least one individual fuel type.
\item included an excessively high proportion ($>$15\%) of respondents who either reported that they cook with an unlisted fuel, that they do not cook at all, or who failed to respond.
\item were flagged in the database as unsuitable for modelling.
\end{itemize}
Surveys which were not included for modelling are shown as black points in the plots of predicted fuel use provided as supplementary material.
\end{appendices}

\bibliographystyle{rss}
\bibliography{library} 
\end{document}